\documentclass[10pt,conference]{IEEEtran}
\IEEEoverridecommandlockouts

\usepackage{cite}
\usepackage{amsmath,amssymb,amsfonts}
\usepackage[linesnumbered,ruled,vlined]{algorithm2e}  
\usepackage{mathtools} 
\usepackage{graphicx}
\usepackage{textcomp}
\usepackage{xcolor}
\usepackage{cuted}
\usepackage{multirow}
\usepackage{amsmath}
\usepackage[english]{babel}
\usepackage{amsthm}
\usepackage{comment}
\usepackage[most]{tcolorbox}
\usepackage{subcaption}
\usepackage{comment}
\usepackage{soul}

\makeatletter
\def\ps@IEEEtitlepagestyle{%
  \def\@oddfoot{\mycopyrightnotice}%
}
\def\mycopyrightnotice{%
  \begin{minipage}{\textwidth}
  \centering \scriptsize
  \copyright 2025 IEEE. Personal use of this material is permitted. Permission from IEEE must be obtained for all other uses, in any current or future media, including reprinting/republishing this material for advertising or promotional purposes, creating new collective works, for resale or redistribution to servers or lists, or reuse of any copyrighted component of this work in other works.
  \end{minipage}
}
\makeatother

\usepackage[left=1.35cm,right=1.35cm,top=1.93cm,bottom=4.3cm]{geometry}

\def\BibTeX{{\rm B\kern-.05em{\sc i\kern-.025em b}\kern-.08em
    T\kern-.1667em\lower.7ex\hbox{E}\kern-.125emX}}
\begin{document}

\title{ CASH: Context-Aware Smart Handover for Reliable UAV Connectivity on Aerial Corridors
\thanks{This research is supported by iSEE-6G project under the Horizon Europe Research and Innovation program with Grant Agreement No. 101139291.
}
}

\author{Abdul Saboor\textsuperscript{1},
         Zhuangzhuang Cui\textsuperscript{1},
         Achiel Colpaert\textsuperscript{1,2},
        Evgenii Vinogradov\textsuperscript{1, 3},
        Sofie Pollin\textsuperscript{1,2}
\\\textsuperscript{1}WaveCoRE of the Department of Electrical Engineering (ESAT), KU Leuven, Leuven, Belgium
\\\textsuperscript{2}imec, Kapeldreef 75, 3001 Leuven, Belgium
\\\textsuperscript{3}NaNoNetworking Center in Catalonia (N3Cat), Universitat Polit\`{e}cnica de Catalunya, Spain
\\Email:\{abdul.saboor, zhuangzhuang.cui,  achiel.colpaert, sofie.pollin\}@kuleuven.be, evgenii.vinogradov@upc.edu}

\maketitle

\begin{abstract}
Urban Air Mobility (UAM) envisions aerial corridors for Unmanned Aerial Vehicles (UAVs) to reduce ground traffic congestion by supporting 3D mobility, such as air taxis. A key challenge in these high-mobility aerial corridors is ensuring reliable connectivity, where frequent handovers can degrade network performance. To resolve this, we present a Context-Aware Smart Handover (CASH) protocol that uses a forward-looking scoring mechanism based on UAV trajectory to make proactive handover decisions. We evaluate the performance of the proposed CASH against existing handover protocols in a custom-built simulator. Results show that CASH reduces handover frequency by up to 78\% while maintaining low outage probability. We then investigate the impact of base station density and safety margin on handover performance, where their optimal setups are empirically obtained to ensure reliable UAM communication.

\end{abstract}


\section{Introduction}
In recent years, urban areas have experienced a huge rise in traffic congestion due to increasing population and vehicles \cite{10718279}. In response, future transportation brings promising alternatives, such as Advanced Air Mobility (AAM) and Urban Air Mobility (UAM), which envision using Unmanned Aerial Vehicles (UAVs) or small drones for tasks including cargo delivery, emergency response, and aerial taxis \cite{cohen2021urban, 10793277}. Therefore, various companies are investing in UAM. For example, Archer Aviation revealed plans to introduce an electric air taxi network in New York City to connect Manhattan with major airports \cite{archer2025nyc}. Similarly, the National Aeronautics and Space Administration’s (NASA) Revolutionary Vertical Lift Technology (RVLT) project is developing UAM aircraft designs for emerging aviation markets \cite{johnson2022nasa}.

There is a need to establish dedicated aerial corridors to facilitate the safe and efficient integration of UAVs into national airspace above cities. An aerial corridor consists of pre-defined flight routes acting as sky highways \cite{10594734}. Similar to Air Traffic Service (ATS) routes, dedicated aerial corridors provide structured UAV routes that reduce collision risks, simplify air traffic management, and enable reliable network planning for seamless connectivity. UAVs flying along the corridors must maintain continuous communication with Ground Base Stations (GBSs) to support command-and-control, telemetry, and data-intensive applications. However, the mobility of UAVs results in frequent transitions between cells, especially in dense urban environments where buildings have a high probability of obstructing Line-of-Sight (LoS) links, leading to frequent handovers \cite{10746345}.

Frequent handovers can degrade the performance of mobile networks by increasing signaling overhead and introducing interruptions \cite{sonmez2024handover}. In contrast, delayed handovers increase the chances of outages or Radio Link Failure (RLF). The mobility of UAVs and blockage in urban environments cause sudden variations in signal quality, leading to unnecessary or late handovers. Thus, optimizing handover decisions is crucial to ensure seamless connectivity, minimize network overhead, and balance mobility support and communication reliability \cite{ullah2023survey}.

Traditionally, 5G New Radio (NR) supports multiple handover events (A1–A5, B1–B2) and Conditional Handovers (CHO). Among these, the A3 event remains the most widely used, which triggers handovers based on Reference Signal Received Power (RSRP), fixed Hysteresis Margin ($\Delta$), and predefined Time-To-Trigger (TTT) intervals. Although A3 is simple, it often results in redundant handovers and network overhead, thus limiting reliability in urban air corridors. 


\begin{figure*}[!t]
\centerline{\includegraphics[width=0.7\linewidth]{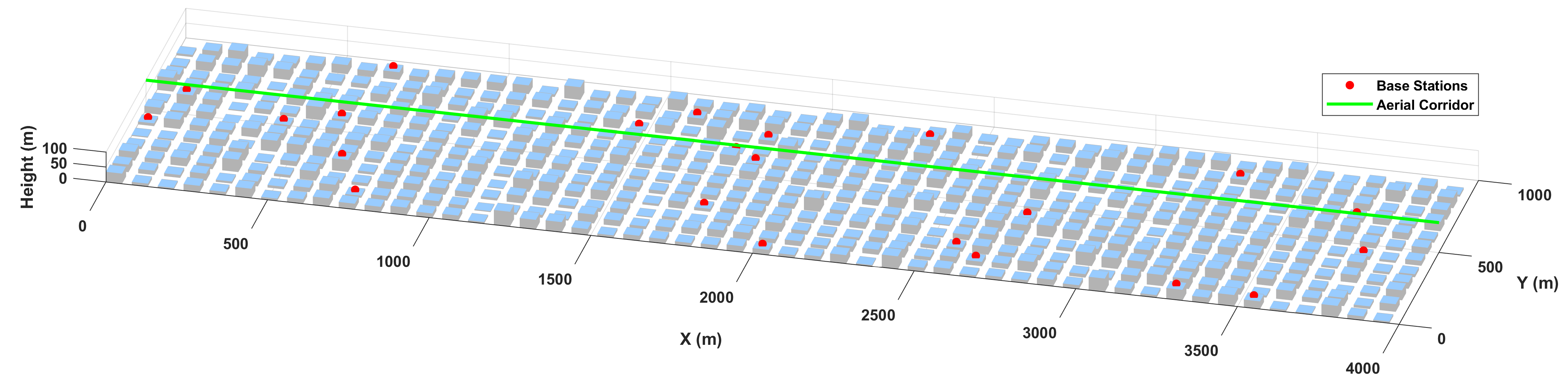}} 
\caption{Top view of simulated urban layout with aerial corridor, buildings, and GBS for handover evaluation.}
\label{Simulator}
\end{figure*}

Several studies have investigated intelligent handover techniques to minimize unnecessary handovers. For example, Mahamod et al. \cite{Mahamod2024HandoverPO} investigate the handover performance using different $\Delta$ and TTT values in a mobility scenario. In contrast, the authors in \cite{10649819} auto-tune the handover measurement and decision-making stage to enhance handover performance. Furthermore, fuzzy-inspired and self-optimizing hysteresis-based analytical techniques (SOHT) offer adaptive handover management to enhance handover performance compared to the traditional A3 method \cite{fuzzy, SOHT}. However, there is still potential to further reduce unnecessary handovers and enhance reliability, particularly under the dynamic and complex conditions of UAM operations.

Although 3GPP NR and NR-Non-Terrestrial Networks (NTN) introduce various handover mechanisms and triggers, including support for aerial platforms, no existing handover protocol has been explicitly proposed in the literature for aerial corridors within UAM scenarios. To bridge these gaps, this paper first introduces the A3T method, which refines the traditional A3 protocol by introducing additional threshold-based handover decisions. After that, it presents the Context-Aware Smart Handover (CASH) protocol, a predictive method that uses the UAV trajectory to select the most suitable serving GBS for handover optimization. Our main contributions are summarized as follows:

\begin{itemize}
    \item We propose the CASH protocol, which integrates trajectory information to select the most suitable serving GBS to minimize unnecessary handovers and outages.
    \item We develop a simulation framework that incorporates various urban environments, aerial corridors, and UAV mobility to evaluate the performance of CASH protocols against existing protocols (A3, fuzzy, and SOHT).
    \item We investigate the optimal density of GBSs required in dense urban environments to balance the number of handovers and outage probability.   
\end{itemize}

The paper is organized as follows: Section~II presents the system model and simulation setup. Section~III details the proposed CASH. Section~IV discusses the simulation results and provides key takeaways. Finally, Section~V concludes the paper.

\section{System Model and Simulation Framework}
We develop a MATLAB-based simulator to evaluate handover performance across diverse urban environments. The framework includes different urban environments generation, three-dimensional (3D) LoS Probability ($P_{\mathrm{LoS}}$) and  Path Loss ($PL$) modeling, base station deployment, UAV mobility, and handover evaluation. We detail the system model as follows. 

\subsection{Urban Layout and Building Modeling}
In this paper, we extend the simulator from our previous studies \cite{saboorEUCAP, saboorWCNC}, featuring a flexible simulation area (with a $4\,\mathrm{km} \times 1\,\mathrm{km}$ area, $A_{\text{land}} = 4\,\mathrm{km}^2$ used in this study). The framework adopts the International Telecommunication Union (ITU)--based Manhattan grid structure \cite{ITU}, capable of generating a city based on three built-up parameters:

\begin{itemize}
    \item $\alpha$: Urban density ratio (building area to land area),
    \item \(\beta\): Building density (buildings/km²),
    \item \(\gamma\): Rayleigh parameter for random building heights.
\end{itemize}

We can generate different urban environments using the combination of built-up parameters. For example, a suburban environment uses $(\alpha, \beta, \gamma) = (0.1, 750, 8)$, urban uses $(0.3, 500, 15)$, dense urban areas adopt $(0.5, 300, 20)$, while urban high-rise environments use $(0.5, 300, 50)$.

Given $\alpha$ and $\beta$, the width $W$ of each square building and the spacing $S$, also called Streets, between adjacent buildings are calculated using $W = 1000 \sqrt{\frac{\alpha}{\beta}}, \quad S = \frac{1000}{\sqrt{\beta}} - W$. The total number of buildings is computed as $N_b = \beta \cdot A_{\text{land}}$, and the height of each building is obtained from $h \sim \text{Rayleigh}(\gamma)$. Fig.~\ref{Simulator} shows a top view of a dense urban environment generated in our simulations.

\subsection{GBS Deployment}

We deploy $n_{\text{GBS}}$, denoted by $\{b_1, b_2, b_3, \dots, b_{n_{\text{GBS}}}\}$, each positioned on a randomly selected rooftop with the main objective to serve aerial traffic. The 3D coordinates of GBS $b_i$ are given by:

\begin{equation}
\mathbf{g}_i = (x_i, y_i, h_i + h_{\text{ext}}), \quad i = 1, 2, \dots, n_{\text{GBS}}
\end{equation}

where $(x_i, y_i)$ represents the center of the roof, $h_i$ is the height of the $i$-th building, and $h_{\text{ext}} = 5$~m accounts for the typical extension of the external antenna above the roof. This height ensures reliable LoS coverage for UAVs flying over the aerial corridor while minimizing the impact of rooftop obstacles such as water tanks, solar panels, or other mechanical equipment. Each GBS transmits at a fixed power $P_{\mathrm{tx}}$
, allowing them to provide a stable connection for UAVs flying over the predefined aerial corridor.

\subsection{Aerial Corridor and UAV Mobility}

To evaluate handover performance, we define an aerial corridor as a straight path from $p_{\text{start}} = (0, Y/2, h_u), \quad p_{\text{end}} = (D, Y/2, h_u)$,
maintaining a fixed altitude $h_u\,$m, constant speed $v_u$~km/h on a trajectory length $D$. This scenario mimics UAM designs where predefined aerial corridors ensure safe and regulated UAV navigation above city infrastructure. The primary reason for using a straight aerial corridor is to simplify navigation, ensure predictable and safe drone movement, and improve communication reliability. It also reduces energy and travel time, making it ideal for efficient urban drone operations. Nevertheless, the proposed simulator can generate different corridor designs, which will be explored in the future.  

If the UAV is moving with $v_u$~km/h at $h_u$, its 3D position at any time $t$ is given in the equation \eqref{eq3}. 

\begin{equation}
\label{eq3} 
\mathbf{u}(t) = (x(t), y(t), h_u), \quad 0 \le t \le T
\end{equation}

In this particular simulation, $x(t)$ evolves linearly from 0 to the trajectory length $D$. The UAV trajectory is discretized into $N_{\text{wp}}$ waypoints for stepwise evaluation of signal strength and handover decisions. The spacing between consecutive waypoints is calculated using equation \eqref{eq4}. 

\begin{equation}
\label{eq4} 
\Delta d = \frac{v_u}{3.6} \cdot \frac{\mathrm{TTT}}{1000}.
\end{equation}

where $v_u$ is the UAV speed in km/h and $TTT$ is the Time-To-Trigger in milliseconds. Given a total trajectory length $D$, the number of waypoints is computed using $N_{\text{wp}} = \left\lceil \frac{D}{\Delta d} \right\rceil$.

\subsection{Wireless Channel and Path Loss Modeling}

At each UAV waypoint \(\mathbf{wp}_t\), the wireless channel between the UAV and each GBS $i$ is characterized by the 3D Euclidean distance $d_i(t) = \left\| \mathbf{wp}_t - \mathbf{g}_i \right\|_2$.

The $P_{\mathrm{LoS}}$ between the UAV and a GBS is determined using a geometric pathloss model considering obstruction from the bounding boxes of buildings. A ray is cast from a particular $b_i$ toward the UAV at $\mathbf{wp}_t$, which checks the ray intersection against the bounding boxes of all buildings between GBS and UAV. If the ray intersects no buildings, the link is classified as LoS; otherwise, it is classified as Non-Line of Sight (NLoS). The $PL$ at 28~GHz frequency $f_c$ is then modeled using \eqref{eq7}, following the approach in \cite{6834753}.

\begin{equation}
\label{eq7} 
PL_i(t) = 
\begin{cases}
61.4 + 20 \log_{10}(d_i(t)), & \text{if LoS} \\
72 +  10n\log_{10}(d_i(t)), & \text{if NLoS}
\end{cases}.
\end{equation}

This model represents typical urban millimeter-wave (mmWave) propagation, where LoS links experience free-space attenuation. In contrast, NLoS links suffer additional diffraction and scattering losses, expressed via $PL$ exponent $n$, ranging from 2.5$-$4, depending on the urban environment. The Reference Signal Received Power (RSRP) from GBS $b_i$ at time $t$ is computed as:

\begin{equation}
\text{RSRP}_i(t) = P_{\mathrm{tx}} + G_{\mathrm{tx}} + G_{\mathrm{rx}} - PL_i(t).
\end{equation}

where $P_{\mathrm{tx}} = 30$~dBm is the UAV's transmit power. $G_{\mathrm{tx}}$ and $G_{\mathrm{rx}}$ are antenna gains, which are not included in the RSRP calculation, as omnidirectional antennas are assumed at both the UAV and GBS. These RSRP values are used to assess link quality and drive handover decisions under various strategies, which are discussed in the following section. 

\section{Handover Strategies}
This section presents five handover strategies evaluated in this study. All schemes are built upon the previously described RSRP-based handover framework but differ in handover triggering conditions, including the hysteresis margin $\Delta$ and TTT. Among them, the A3T and CASH strategies are proposed in this work, while A3, fuzzy-inspired, and SOHT are adapted from existing literature.

\subsection{A3 Handover}
The standard A3 handover in 5G NR is triggered when the RSRP of a neighboring GBS $\text{R}_{b_c}$ becomes higher than the RSRP of serving GBS $\text{R}_{b_s}$ by a fixed hysteresis margin $\Delta$, sustained over the TTT period, as given in equation \eqref{a3eq}. 

\begin{equation}
\label{a3eq}
\text{R}_{b_c}(t) > \text{R}_{b_s}(t) + \Delta, \quad \forall t \in [t_0, t_0 + \text{TTT}].
\end{equation}

The UAV sends a measurement report if this condition holds for the entire TTT duration. Upon approval, the serving GBS issues a handover command, and the process finalizes once the UAV attaches to the new GBS. Despite its simplicity, static $\Delta$ and TTT values make the A3 approach vulnerable to ping-pong effects or delayed switching in dynamic UAV environments.

\subsection{SOHT: Self-optimizing Hysteresis and TTT}
SOHT works on the same principle as the A3 handover procedure in equation \eqref{a3eq}. However, unlike A3 with static $\Delta(t)$ and TTT, SOHT dynamically adjusts these values based on the UAV's speed and inter-site distance. These adaptive parameters are computed analytically as:

\begin{align}
\Delta(t) &= \left(10n \log_{10}\left(\frac{2v_u \psi \sin |\phi_t|}{d_t} + 1\right)\right)^{-1}. \\
\mathrm{TTT}(t) &= \left(\frac{d_t}{v_u \sin |\phi_t|} \left(\frac{\sqrt{2v_u \psi \sin |\phi_t| + d_t}}{d_t} - 1\right)\right)^{-1}.
\end{align}

where $v_u$ is the UAV speed, $d_t$ is the separation from the BS, $\phi_t$ is the bearing angle, $\psi$ is the measurement periodicity, and $n$ is the PL exponent. These expressions help in proactive and stable handover decisions according to user mobility, resulting in improved handover performance.

\subsection{Fuzzy-Inspired Handover}
This method employs a Fuzzy Logic Controller (FLC) to dynamically tune $\Delta$ and TTT based on $v_u$, $\mathrm{R}_{b_s}(t)$, and estimated GBS load $\lambda$. These inputs are fuzzified into linguistic categories (``slow,'' ``weak,'' ``high load'') and processed through a fuzzy inference engine. The $\Delta$ is obtained using $\Delta = \mathrm{FLC}(v_u, \mathrm{R}_{b_s}(t), \lambda)$, and TTT is computed via a weighted function combining normalized input effects: 


\begin{equation}
\mathrm{TTT} = w_1 f_{\mathrm{RSRP}} + w_2 f_{\mathrm{load}} + w_3 f_{\mathrm{v}}.
\end{equation}

Here, $w_1$, $w_2$, and $w_3$ are weights derived from the input values, and each $f(\cdot)$ term estimates the relative impact of its associated metric. This hybrid approach minimizes RLF and ping-pong effects under varying mobility and load conditions.

\subsection{A3T: A3 Threshold Handover}

The proposed A3T extends the standard A3 protocol by introducing a signal quality check using the RSRP threshold $\tau_{\min}$. Therefore, the handover condition becomes:

\begin{equation}
\label{a3teq}
\text{R}_{b_c}(t) > \text{R}_{b_s}(t) + \Delta, \: \text{\&} \: \text{R}_{b_s}(t) \leq \tau_{\min} + \delta,  \forall t \in [t_0, t_0 + \mathrm{TTT}].
\end{equation}

A3T ensures that handovers occur when the new link is genuinely stronger and the current one is weak, reducing unnecessary handovers. Therefore, it is effective in dynamic environments like UAV networks, where link quality fluctuates rapidly, leading to unnecessary handovers.

While all these schemes try to optimize handover-triggering events based on parameters like $\Delta$, TTT, or by introducing a threshold, no existing scheme considers the most suitable GBS, which can serve UAVs for longer periods on aerial corridors to avoid unnecessary handovers. Therefore, we present the CASH protocol in the following subsection.

\begin{algorithm}[!t]
\small
\caption{CASH: Context-Aware Smart Handover}
\KwIn{RSRP vector $\mathbf{r}$, UAV position $\mathbf{u}$, GBS positions $\{\mathbf{g}_i\}$, $\Delta$, TTT, $\tau_{\min}$, $\delta$, $y_{\text{AC}}$}
\KwOut{$b_s$, handover count $\mathcal{H}$, outage count $\mathcal{\hat{O}}$}

\If{first time step}{
    Initialize $b_s$ $\leftarrow$ \textsc{argmax}($\mathbf{r}$)\;
    Initialize \texttt{ttt\_counter} $\leftarrow 0$\;
}

Compute candidates: $i \in \mathcal{C}$ such that $r_i \geq \tau_{\min}$ and $x_i > x_{\text{UAV}}$\;

\If{$\mathcal{C} = \emptyset$}{
    Reset \texttt{ttt\_counter}\;
}
\Else{
    Compute score for each $i \in \mathcal{C}$:
    \[
    \text{score}_i = \frac{x_i - x_{\text{UAV}}}{1 + |y_i - y_{\text{AC}}|}
    \]
    Select best GBS $b_{c}$ $\leftarrow \arg\max_{i \in \mathcal{C}} \text{score}_i$\;

    \If{$r_{b_{c}} > r_{b_s} + \Delta$ and $r_{b_{s}} \leq \tau_{\min} + \delta$}{
        Increment \texttt{ttt\_counter}\;
        \If{\texttt{ttt\_counter} $\geq TTT$}{
            Perform handover: $b_s$ $\leftarrow$ $b_{c}$\;
            Increment $\mathcal{H}$\;
            Reset \texttt{ttt\_counter}\;
        }
    }
    \Else{
        Reset \texttt{ttt\_counter}\;
    }
}

\If{$r_{\text{current\_gbs}} < \tau_{\min}$ dBm}{
    Increment $\hat{\mathcal{O}}$\;
}
\end{algorithm}

\subsection{CASH: Context-Aware Smart Handover (Proposed)}
The proposed CASH protocol uses the same handover trigger criteria as in \eqref{a3teq}. However, it predicts the most suitable neighboring GBS by filtering candidates ahead of the UAV’s trajectory and ranking them using a geometric score that favors forward-aligned and corridor-centered base stations, as given in Algorithm 1. From this filtered set, it selects the BS with the highest geometric score, calculated by \eqref{cashscore}.

\begin{equation}
\label{cashscore}
\text{score}_i = \frac{x_i - x_{\text{UAV}}}{1 + |y_i - y_{\text{AC}}|}.
\end{equation}

This score favors the furthest GBSs ahead of the UAV along its flight path (large $x_i - x_{\text{UAV}}$), which are positioned close to the aerial corridor centerline ($ y_{\text{AC}}$). The smaller $y_{\text{AC}}$, also called lateral offset, increases the chances of a LoS link between UAV-GBS, as GBS aligned with the corridor are less likely to be blocked by surrounding buildings. The additive constant in the denominator ensures numerical stability and penalizes large lateral offsets, effectively discouraging the selection of GBSs far off to the sides. As a result, the score favors handovers to GBSs with strong geometry and higher LoS probability. A higher score indicates a more reliable future link, and a handover is triggered if the candidate’s RSRP exceeds the current serving GBS by a margin $\Delta$ and remains below a reliability threshold $\tau_{\min} + \delta$. Here, $\delta$ is the Handover Safety Margin (HSM), introduced to ensure that handovers are initiated early enough to prevent outages $\hat{\mathcal{O}}$ or RLF. In general, CASH preserves the $O(n_{\text{GBS}})$ complexity of A3. In addition to the standard RSRP checks, CASH filters neighboring GBSs ahead of the UAV ($O(n_{\text{GBS}})$) and, for the remaining candidates $C$, computes a geometric score and selects the best option ($O(|C|) \leq O(n_{\text{GBS}})$). These lightweight operations introduce only minimal overhead compared to baseline A3.


\section{Results and Discussion}
We implement a MATLAB simulator (see Section II) to benchmark the performance of the proposed and baseline handover strategies. All simulations were executed in MATLAB R2024a on a server with an Intel Xeon Gold 5220 CPU (30 parallel threads) and 256\,GB RAM, ensuring reproducibility. Each scenario involves a UAV flying in an aerial corridor over a pre-defined urban environment. GBSs are deployed on rooftops using uniform random sampling. At each time step, the UAV evaluates the handover condition based on the defined strategy (i.e., CASH, A3). RSRP values are computed for all GBSs using LoS/NLoS-aware 3D PL models. The simulation parameters are given in Table \ref{simparams}. For each result, we simulate 500 unique cities to obtain an average.

\begin{table}[t]
\caption{Simulation Parameters}
\centering
\begin{tabular}{|p{1.5cm}|p{1.5cm}|p{2cm}|p{2cm}|}
\hline
\textbf{Parameter} & \textbf{Value} & \textbf{Parameter} & \textbf{Value} \\
\hline
$\alpha$ & 0.1–0.5 & $v_u$ & 100 km/h \\ \hline
$\beta$ & 300–750/km$^2$ & $h_u$ & 100 m \\ \hline
$\gamma$ & 8–50~m & TTT & 100-300~ms \\ \hline
$h_{\text{ext}}$ & 5 m & $f_c$ & 28~GHz \\ \hline
$P_{\mathrm{tx}}$ & 30 dBm & $\tau_{\min}$ & $-101.5$ dBm \\ \hline
$n_{\text{GBS}}$ & 2-10/km$^2$ & $n$ & 2.5-4 \\ \hline
\end{tabular}
\label{simparams}
\vspace{-.5cm}
\end{table}

\begin{figure}[!t]
    \centering
    \begin{subfigure}[t]{0.49\linewidth}
        \centering
        \includegraphics[width=\linewidth]{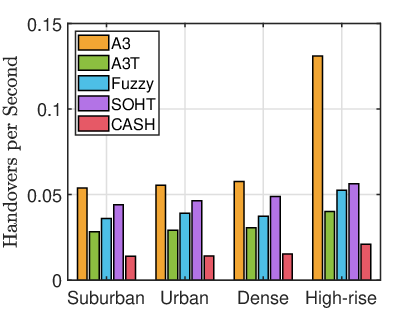}
        \caption{Handover frequency}
        \label{HF4}
    \end{subfigure}%
    \hfill
    \begin{subfigure}[t]{0.49\linewidth}
        \centering
        \includegraphics[width=\linewidth]{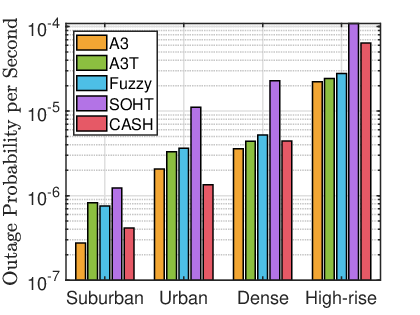}
        \caption{Outage probability}
        \label{OP4}
    \end{subfigure}
    \caption{Handovers and $OP$ for different handover strategies across varying urban environments ($n_{\rm GBS}$ = 6/km$^2$).}
    \label{HOP4}
\end{figure}

Fig. \ref{HOP4} illustrates the handover frequency (handover per second) and Outage Probability ($OP$) for different handover strategies across varying urban environments with 6~GBS/km$^2$. We use 6~GBS/km$^2$ as a balanced case to show how each method performs under typical deployment conditions by avoiding extreme cases.  The $OP$ is the percentage of time the UAV has no viable GBS with RSRP $\geq \tau_{\min}$. The standard A3 follows a conservative approach of selecting the best GBS whenever available, leading to frequent handovers, particularly in high-rise environments with a high probability of blockages. However, this helps maintain a lower $OP$ compared to the other protocols. In general, outages and handovers increase in all protocols with an increase in urban density due to frequent blockages and rapid channel dynamics. However, A3T, fuzzy, and SOHT mitigate this moderately by tuning $\Delta$ or TTT. In contrast, CASH proactively anticipates the most suitable GBS in an urban environment using trajectory-aware scoring and an HSM, which helps it to significantly reduce handovers and prevent RLFs, even under high-rise conditions.

\begin{tcolorbox}[colback=gray!10, colframe=black!30, boxrule=0.4pt, sharp corners, left=2pt, right=2pt, top=2pt, bottom=2pt]
\footnotesize \textbf{Takeaway 1:} On average across all environments, CASH reduces handover frequency by $\approx$78\% compared to traditional A3, while maintaining comparable outage performance for $n_{\rm GBS} = 6$. 
\end{tcolorbox}

\begin{figure}[!t]
    \centering
    \begin{subfigure}[t]{0.49\linewidth}
        \centering
        \includegraphics[width=\linewidth]{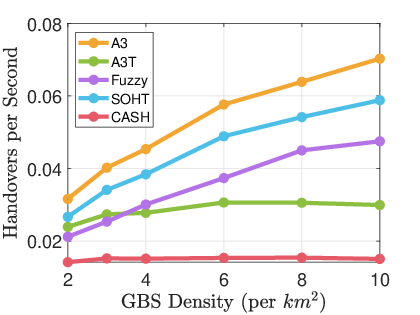}
        \caption{Handover frequency}
        \label{HF}
    \end{subfigure}%
    \hfill
    \begin{subfigure}[t]{0.49\linewidth}
        \centering
        \includegraphics[width=\linewidth]{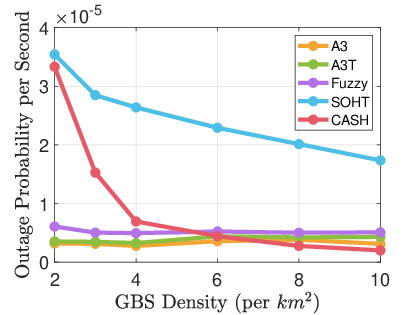}
        \caption{Outage probability}
        \label{OP}
    \end{subfigure}
    \caption{Impact of GBS density on handover frequency and outage probability in dense urban environment. }
    \label{DU_HvsOP}
\end{figure}

Fig. \ref{DU_HvsOP} investigates GBS density's impact on handover frequency and $OP$ in the most challenging dense-urban environments.
 We can observe from Fig. \ref{HF} that with the increases in GBS density, A3 and SOHT have a sharp rise in handovers due to their reactive behavior of choosing the best-serving GBS. A3T and fuzzy improve stability modestly, while CASH consistently maintains the lowest handover rate by proactively avoiding short-lived connections.

\begin{figure}[!t]
    \centering
    \includegraphics[width=0.75\linewidth]{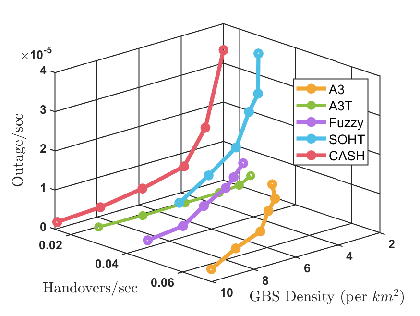}
    \caption{3D trade-off between handovers and $OP$ across varying GBS densities in dense urban environment.}
    \label{3Dview}
\end{figure}

In terms of reliability, GBS density generally helps reduce outages for most methods until a saturation point. SOHT shows high outages at low densities due to late triggering, whereas A3, A3T, and fuzzy stabilize quickly. CASH demonstrates a steep drop in $OP$ as density increases from 2 to 4 GBS/km$^2$. After 6 GBS/km$^2$, we observe that CASH outperforms the most conservative A3 in $OP$ because the higher density of GBS increases the chances of finding GBSs that are better aligned with the aerial corridor and offer higher $P_{\mathrm{LoS}}$, allowing longer and reliable connections. Lastly, Fig.~\ref{3Dview} visualizes the trade-off between handovers and outages across GBS densities. CASH consistently occupies the bottom-left region, highlighting its ability to balance reliability and signaling overhead (handovers) better than other methods.


\begin{tcolorbox}[colback=gray!10, colframe=black!30, boxrule=0.4pt, sharp corners, left=2pt, right=2pt, top=2pt, bottom=2pt]
\footnotesize 
\textbf{Takeaway 2:} After a certain point (e.g., $>$6 GBS/km$^2$), additional GBSs do not significantly improve reliability for existing protocols (A3, Fuzzy). Instead, they increase the risk of unnecessary handovers. However, CASH filters these using its score, maintaining low outages and handovers.  \\
\textbf{Takeaway 3:} Outages increase due to limited GBS options at low densities, whereas unnecessary handovers increase without added reliability at high GBS densities. For dense urban aerial corridors, a GBS density of 4-6/km$^2$ offers the best performance against outage, handover, and GBS density.
\end{tcolorbox}

\begin{figure}[!t]
    \centering
    \includegraphics[width=0.75\linewidth]{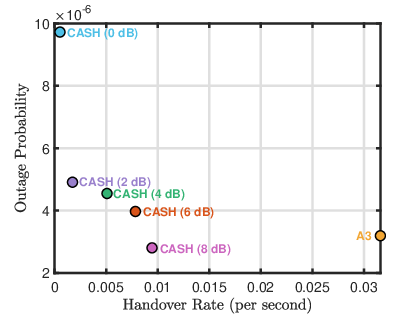}
    \caption{Impact of $\delta$ tuning on CASH performance. Higher $\delta$ lowers outage but increases handover.}
    \label{HSM}
\end{figure}

Fig. \ref{HSM} addresses the 
higher outage observed for CASH at lower GBS density (e.g., 2/km$^2$) in Fig. \ref{3Dview}. It illustrates how tuning the safety margin $\delta$ in CASH affects performance. With the increase in $\delta$, CASH adopts a more conservative approach by initiating handovers earlier, resulting in more handovers but fewer outages and vice versa. In this study, $\delta$ is optimized empirically; however, its optimal value depends on the GBS density and urban layout. In the future, our goal is to develop an adaptive scheme that dynamically tunes $\delta$ based on the real-time network and environmental conditions.


\begin{tcolorbox}[colback=gray!10, colframe=black!30, boxrule=0.4pt, sharp corners, left=2pt, right=2pt, top=2pt, bottom=2pt]
\footnotesize 
\textbf{Takeaway 4:} The $\delta$ in CASH can be tuned to balance reliability and handover stability under sparse GBS conditions. However, dynamic adaptation of $\delta$ (optimization) based on environment and GBS density is required. 
\end{tcolorbox}

\section{Conclusion}

This paper introduced the CASH protocol for UAV communication in aerial corridors. It proactively selects the most stable GBS using a geometric scoring function and margin $\delta$ to minimize unnecessary handovers, thus minimizing communication overhead. The simulation results in diverse urban environments and GBS densities demonstrate that CASH significantly reduces the handover frequency, up to 78\% compared to A3, while maintaining a low probability of outage. We also show that tuning the HSM parameter $\delta$ allows CASH to adapt to sparse deployments of GBS. The main limitation of CASH lies in its reliance on predefined corridors and accurate trajectory information, which may reduce effectiveness in free-flight/corridor-free scenarios. Future work will explore its adaptability to dynamic trajectories, environment-aware HSM tuning, and varying flight altitudes.


\bibliographystyle{IEEEtran}
\bibliography{ref}

\end{document}